\documentclass[iop]{emulateapj}

\usepackage{graphicx,amssymb}
\usepackage[usenames,dvips]{color}

\newcommand{\kms}{\ifmmode {\rm km\,s}^{-1} \else km\,s$^{-1}$\fi}

\shorttitle{Methanol Masers in Supernova Remnants}

\begin{document}

\title{Detection of Class I Methanol (CH$_3$OH) Maser Candidates in
  Supernova Remnants}

\author{Y.~M.~Pihlstr\"om\altaffilmark{1}}\affil{Department of Physics
  and Astronomy, University of New Mexico, MSC07 4220, Albuquerque, NM
  87131}

\email{ylva@unm.edu}

\altaffiltext{1}{Y.~M.~Pihlstr\"om is also an Adjunct Astronomer at the
 National Radio Astronomy Observatory}

\author{L.~O.~Sjouwerman, D.~A.~Frail, M.~J.~Claussen}
\affil{National Radio Astronomy Observatory, P.O. Box 0, Lopezville
 Rd.\ 1001, Socorro, NM 87801}

\and

\author{R.~A.~Mesler, B.C.~McEwen} \affil{Department of Physics
 and Astronomy, University of New Mexico, MSC07 4220, Albuquerque, NM
87131}

\begin{abstract}
  We have used the Karl G.\ Jansky Very Large Array (VLA) to search
  for 36 GHz and 44 GHz methanol (CH$_3$OH) lines in a sample of 21
  Galactic supernova remnants (SNRs). Mainly the regions of the SNRs
  with 1720 MHz OH masers were observed. Despite the limited spatial
  extent covered in our search, methanol masers were detected in both
  G1.4$-$0.1 and W\,28. Additional masers were found in
  Sgr\,A\,East. More than 40 masers were found in G1.4$-$0.1 which we
  deduce are due to interactions between the SNR and at least two
  separate molecular clouds. The six masers in W\,28 are associated
  with the molecular cloud that is also associated with the OH maser
  excitation. We discuss the possibility that the methanol maser may
  be more numerous in SNRs than the OH maser, but harder to detect due
  to observational constraints.
\end{abstract}


\section{Introduction}
\label{intro}

In the last few years Class I methanol (CH$_3$OH) masers have been
associated with supernova remnants (SNRs). The new capabilities of the
Karl G.\ Jansky Very Large Array (VLA) enabled the detection of 36 GHz
methanol masers in the Sgr\,A\,East molecular cloud/SNR interaction
region \citep{sjouwerman10}. In the same region, 44 GHz methanol
masers have been found \citep{yusef-zadeh08,pihlstrom11}. These
transitions are pumped by collisions, similar to 1720 MHz OH
masers. The 1720 MHz OH maser transition has been recognized as an
important molecular tracer of shocks in SNRs \citep{wardle02} and,
through extensive observations and theory, became a signpost for the
interaction of SNRs with molecular clouds
\citep{claussen97,green97,koralesky98,brogan00,hewitt08}. The OH
molecule is formed downstream of a compression-type shock that has
propagated into a molecular cloud (at 20$-$30 \kms) with densities of
order $n_0 = 10^4 - 10^5$ cm$^{-3}$ at temperatures of 30$-$120
K. Under these conditions a strong collisionally pumped maser
  transition results for the OH molecule at 1720 MHz
  \citep{elitzur76,lockett99,wardle99,hewitt09}. The values of the
  physical conditions predicted have been verified by various
  multiwavelength observations of SNRs harboring OH 1720 MHz masers
  \citep{frail98,reach05,lazendic10,brogan13}. Radiative pumping of
  the 1720 MHz maser is also possible, in which case it is usually
  accompanied by bright 1665 MHz and 1667 MHz emission, as observed in
  star forming regions and evolved stars
  \citep{cragg02,karlsson03}. A detection of a single 1720 MHz OH
maser, without a trace of emission at other OH hyperfine
transitions (at 1612, 1665 and 1667 MHz) is however unambiguous proof
of an interaction, unlike other tracers such as HI or CO emission.
From previous VLA observations it was concluded that methanol masers
in Sgr\,A\,East are located near, but not perfectly co-spatial with
the 1720 MHz OH masers indicating that they trace different shocks or
different regions of the shock.

Theoretical modeling of methanol shows that the brightest
shock-excited masers are expected for the transitions at 36.169 GHz
and at 44.070 GHz, with weaker masers at the 84.521 GHz and 95.169 GHz
transitions \citep{morimoto85,cragg92}.  As can be expected, the
  collisionally pumped 36 GHz and 44 GHz maser lines of methanol have
  been found associated with outflows in star forming regions
  \citep[e.g.,][]{kalenskii10,fish11}. Further work on star forming
  regions suggest that these methanol maser transitions can be excited
  over a larger range of densities and temperatures than 1720 MHz OH.
  Statistical equilibrium calculations supported by large velocity
  gradient modelling of the maser spectra in specific sources estimate
  parameter ranges of $n=10^4-10^7$ cm$^{-3}$ and $T=30-200$ K
  \citep{leurini07,pratap08}. \citet{litovchenko12} further reports on
  a 35\% detection rate of 1720 MHz OH masers in sample of sources
  harboring collisionally pumped methanol masers. Compared to the 1720
  MHz OH which is only detected in 10\% of all Galactic SNRs, the
  expanded range of physical conditions allowing methanol maser action
  potentially makes methanol a new and more widespread signpost for
  SNR-molecular cloud interactions. If methanol masers would be
detected in a higher percentage of SNRs than OH masers, they open up
the possibility to, e.g., further constrain models of hadronic cosmic
ray acceleration in SNRs. Such models depend strongly on distance
  and density
  \citep[e.g.,][]{drury94,kelner06,cristofari13}. Molecular lines
like OH and methanol are excited under certain conditions and density
ranges, yielding density estimates when detected. The maser
  velocities often agree very accurately with the SNR systemic
  velocity \citep{larionov07}, allowing the line velocities to be
used for kinematic distance estimates, and possibly in the longer term
for even more accurate distance measurements using trigonometric
parallaxes.

This paper reports on the results of using methanol as a possible
SNR/molecular cloud interaction tracer. In Sect.\ \ref{obs} and
\ref{results} we describe the observations and results, respectively,
which then are discussed in Sect.\ \ref{discussion}.

\begin{deluxetable*}{ll|cc|ccclc}[tbh]
  \tabletypesize{\scriptsize} \tablecaption{Targets searched for 36
    GHz and 44~GHz methanol masers} \tablewidth{0pt}
  \tablehead{\colhead{Source} & \colhead{Other name} & \colhead{Right Ascension}\tablenotemark{a} & \colhead{Declination}\tablenotemark{a} & \colhead{Distance}& \colhead{$V_{\rm OH}$} & \colhead{Ptgs} &  \colhead{SNR Size} & \colhead{\% Covered} \\
    & & \multicolumn{2}{c}{(J2000)} & (kpc) & (\kms) & (36 GHz) &
    ($\arcmin\times\arcmin$) & (36 GHz) } \startdata
  G189.1$+0.3$ & IC\,433      & 06 18 02.7 & $+$22 39 36 & 1.5 & $-15$ &  7 & 45$\times$25 & 1.0 \\
  &              &            &             &     & $+20$ &  1 &              & 0.1  \\
  G346.6$-0.2$ &              & 17 10 19.0 & $-$40 10 00 & 11  & $-75$ &  3 &  8$\times$8  & 7.3 \\
  G348.5$+0.1$ & CTB\,37A     & 17 15 26.0 & $-$38 28 00 &  8  & $-65$ &  9 & 15$\times$15 & 6.3 \\
  &              &            &             &     & $+22$ &  1 &              & 0.7  \\
  G349.7$+0.2$ &              & 17 17 59.0 & $-$38 28 00 & 11  & $+16$ &  5 & 2.5$\times$2 & 100 \\
  G357.7$+0.3$ & & 17 38 35.0 & $-$30 44 00 & 6.4 & $-36$ & 4 &
  24$\times$24 & 1.1 \\
  G357.7$-$0.1 & MSH 17-39 & 17 40 29.0 & $-$30 58 00 & 6 & $+12$ & 1
  & 20$\times$20 & 0.4 \\
  G359.1$-0.5$ &              & 17 45 30.0 & $-$29 57 00 & 5   & $-4$  &  6 & 24$\times$24 & 1.6 \\
  G000.0$+0.0$ & Sgr\,A\,East & 17 45 44.0 & $-$29 00 00 & 8.5 & $+35$\tablenotemark{b} &  28 & 3.5$\times$2.5& 100\\
            &              &            &             &     & $+132$& 1  &              & 17.9\\
G001.05$-0.15$&             & 17 48 30.0 & $-$28 09 00 & 8.5 & $-1$  &  1 & 8$\times$8   & 2.4 \\
G001.4$-0.1$ &              & 17 49 39.0 & $-$27 46 00 & 8.5 & $-2$  &  5 & 10$\times$10 & 7.8 \\
G005.7$-0.0$ &              &            &             & 3.2 & $+13$ &  1 & 4$\times$8   & 4.9\\
G005.4$-1.2$ & Milne\,56    & 18 02 10.0 & $-$24 54 00 & 5.2 & $-21$ &  1 & 35$\times$35 & 0.1\\
G006.4$-0.1$ & W\,28        & 18 00 30.0 & $-$23 26 00 & 2   & $+7$  & 29 & 48$\times$48 & 2.0\\
G008.7$-0.1$ & W\,30        & 18 05 30.0 & $-$21 26 00 & 3.9 & $+36$ &  1 & 40$\times$40 & 0.1\\
G009.7$-0.0$ &              & 18 07 22.0 & $-$20 35 00 & 4.7 & $+43$ &  1 & 15$\times$11 & 0.9\\
G016.7$+0.1$ &              & 18 20 56.0 & $-$14 20 00 & 14  & $+20$ &  1 & 4$\times$4   & 9.8\\
G021.8$-0.6$ & Kes\,69      & 18 32 45.0 & $-$10 08 00 & 11  & $+69$ &  1 & 20$\times$20 & 0.4 \\
G031.9$+0.0$ & 3C\,391      & 18 49 25.0 & $-$00 55 00 &  9  &$+108$ &  2 & 7$\times$5   & 8.9\\
G032.8$-0.1$ & Kes\,78      & 18 51 25.0 & $-$00 08 00 & 11  & $-86$ &  1 & 15$\times$15 & 0.7\\
G034.7$-0.4$ & W\,44        & 18 56 00.0 & $+$01 22 00 & 2.5 & $+44$ & 23 & 30$\times$25 & 4.8\\
G049.2$-0.7$ & W\,51C       & 19 23 50.0 & $+$14 06 00 & 6   & $+70$ &  1 & 25$\times$20 & 0.3\\
\enddata
\label{targets}
\tablenotetext{a}{Positions are from Green's Catalogue of Supernova Remnants \citep{green09}}
\tablenotetext{b}{Center frequency shifted slightly at a few scheduling blocks to cover a larger velocity range}
\end{deluxetable*}

\section{Observations}\label{obs}
We report on the results from two VLA programs (10B-146 and S3115).
Our sample was drawn from all SNRs with known 1720~MHz~OH masers, and
therefore known to interact with a molecular cloud \citep{frail11}. Of
those, the 21 listed in Table \ref{targets} have declinations that can
be observed from the VLA.

Data were taken during the fall of 2010 and spring of 2011, with the
VLA Ka- and Q-band systems to observe the 36.169 GHz and the 44.070
GHz lines.  Sgr\,A\,East was only observed in the 44 GHz transition,
as another observing program was already ongoing for the 36 GHz
transition. The array was in C, CnB and B configurations, resulting in
synthesized beam sizes of 0.15\arcsec$\leq\theta\leq$2.4\arcsec\ at 36
GHz and 0.12\arcsec$\leq\theta\leq$2.5\arcsec\ at 44~GHz with typical
synthesized beam sizes of 0.30\arcsec$\times$0.15\arcsec\ and
0.25\arcsec$\times$0.13\arcsec, respectively. The VLA field-of-view at
36 GHz and 44 GHz is 1.25\arcmin\ and 1.02\arcmin. In comparison, all
SNRs have a large angular extent (see Table \ref{targets}), and with
the small primary beams at Ka- and Q-band, for most sources only a
limited part of the SNR angular extent could be covered. The small
field-of-view relative to the SNR size presents a challenge for the
survey. Our solution was to target the majority of the methanol survey
pointings toward the positions of previously known {\it hydroxyl (OH)}
masers. This strategy was used to good effect in our initial methanol
detections in Sgr\,A\,East. A few additional pointings in G1.4$-$0.1
included positions where the radio continuum showed features
(peaks/ridges).  Additional observations in W\,28 pointed toward
shocked gas as indicated by the presence of H$_2$ emission. In total,
84 (or 108) pointings were observed in S3115 and 21 (or 26) pointings
in 10B-146 at 36 GHz (or 44) GHz respectively.

\begin{figure}[thb]
  \label{g1.4a_spec}
  \resizebox{8cm}{!}{\includegraphics{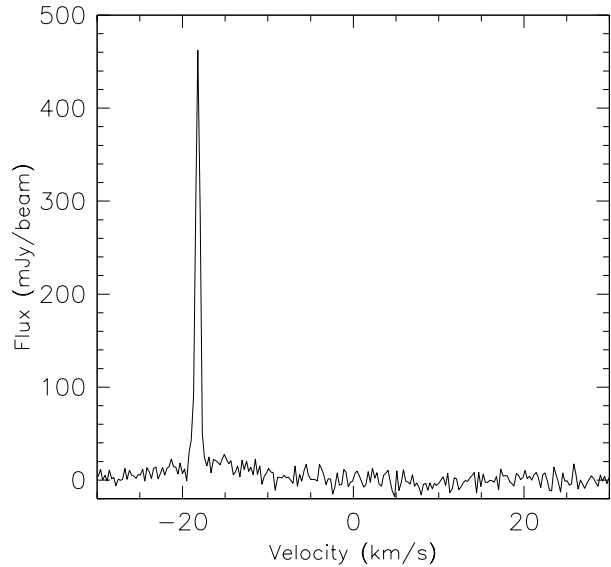}}
  \caption{An example of a maser feature detected in pointing position
    E in G1.4$-$0.1.}
\end{figure}

The data were calibrated using NRAO's Astronomical Image Processing
System (AIPS) and imaged with robust weighting using standard AIPS
procedures. Each observation used 256 frequency channels over 8~MHz
bandwidth, resulting in a channel width of 31.25~kHz or 0.26 and 0.21
\kms\ at 36 GHz and 44~GHz respectively. The bandwidth covered a
velocity range of 67 \kms\ and 55 \kms\ at 36 GHz and 44~GHz
respectively, which was centered at the velocity of the OH masers
($V_{\rm OH}$ in Table \ref{targets}). In some cases two velocity
settings were used, also noted in Table \ref{targets}. For each
pointing position, the on-source integration time was approximately 10
minutes. The precise integration time varied somewhat between the
positions depending on how the observations fit within a 2 hour
scheduling block. Typical final rms noises are between 7 and 10 mJy
per channel, respectively. No continuum emission was found in the
fields.

\begin{deluxetable*}{lll|l|l|rcl}
\tabletypesize{\scriptsize}
\tablecaption{Detected masers}
\tablewidth{0pt}
\tablehead{\colhead{Source} & \colhead{Transition} & \colhead{Pointing} & \multicolumn{2}{c}{Maser position} & \colhead{V$_{CH_3OH}$} & \colhead{$\Delta$V} & \colhead{I$_{peak}$}\\
& (GHz) & &  RA (J2000) & Dec (J2000) & (\kms) & (\kms) & (Jy/beam)\\}
\startdata
G000.0$+0.0$ & 36,44  &   & Multiple & Multiple\tablenotemark{a} & & & \\
G001.4$-0.1$ & 36     & A & 17 49 31.09 & $-$27 47 36.3\tablenotemark{b} & $+27.5$ & 1.6 & 0.09 \\
            & 36     & B & 17 49 38.77 & $-$27 44 22.1 & $+28.8$ & 1.6 & 0.16 \\
            & 36     & B &             &               & $+27.5$ & 0.5 & 0.16 \\
            & 36     & B & 17 49 38.41 & $-$27 44 18.8 & $+23.9$ & 1.0 & 0.28 \\
            & 36     & B & 17 49 38.23 & $-$27 44 19.0 & $+19.0$ & 0.5 & 0.14 \\
            &        &   &             &               & $+17.2$ & 0.3 & 0.23 \\
            & 36     & B & 17 49 38.15 & $-$27 44 16.8 & $+24.9$ & 0.8 & 0.23 \\
            & 36     & B & 17 49 38.09 & $-$27 44 19.9\tablenotemark{c}& $+14.6$ & 1.0 & 0.20 \\
            &        &   &             &               & $+19.0$ & 1.6 & 0.17 \\
            & 36     & B & 17 49 37.99 & $-$27 44 17.5\tablenotemark{c}& $+25.7$ & 1.6 & 0.30 \\
            &        &   &             &               & $+26.5$ &     & 0.24 \\
            & 36     & B & 17 49 37.91 & $-$27 44 23.7\tablenotemark{c}& $+19.5$ & 0.5 & 0.30 \\
            & 36     & B & 17 49 37.85 & $-$27 44 22.0\tablenotemark{c}& $+24.4$ & 1.8 & 0.18 \\
            & 36     & B & 17 49 37.68 & $-$27 44 23.0\tablenotemark{c}& $+17.7$ & 1.0 & 0.16 \\
            & 36     & B & 17 49 37.08 & $-$27 44 18.8 & $+23.1$ & 0.5 & 0.18 \\
            & 36     & B & 17 49 36.82 & $-$27 44 20.6\tablenotemark{c}& $+28.6$ & 0.8 & 0.25 \\
            & 36     & B & 17 49 36.63 & $-$27 44 21.3\tablenotemark{c}& $+20.3$ & 1.8 & 0.33 \\
            &        & B &             &               & $+29.9$ & 0.3 &  0.10 \\
            & 36     & B & 17 49 36.56 & $-$27 44 21.1\tablenotemark{c}& $+20.3$ & 2.3 & 0.21 \\
            &        & B &             &               & $+24.7$ & 0.5 & 0.20  \\
            & 36     & B & 17 49 36.43 & $-$27 44 21.9\tablenotemark{c}& $+18.5$ & 0.5 & 0.89 \\
            &        & B &             &               & $+21.3$ & 1.3 & 0.22 \\
            & 36     & C & 17 49 48.53 & $-$27 43 15.8 & $+16.9$ & 1.6 & 0.09 \\
            & 36     & C & 17 49 48.47 & $-$27 44 31.2 & $-22.7$ & 1.0 & 0.09 \\
            & 36     & C & 17 49 48.23 & $-$27 43 55.0 & $+7.6$ & 1.8 & 0.18 \\
            & 36     & C & 17 49 47.59 & $-$27 44 03.1 & $-21.9$ & 2.3 & 0.07 \\
            & 36     & C & 17 49 47.56 & $-$27 44 10.4\tablenotemark{d}& $-32.6$ & 0.5 & 0.14 \\
            & 36     & C & 17 49 46.98 & $-$27 44 10.4 & $-27.1$ & 0.8 & 0.07 \\
            & 36     & C & 17 49 46.95 & $-$27 44 16.8 & $-29.2$ & 1.3 & 0.08 \\
            & 36     & C & 17 49 45.87 & $-$27 44 11.8 & $-31.3$ & 0.5 & 0.07 \\
            & 36     & C & 17 49 45.43 & $-$27 44 19.4 & $-24.8$ & 0.5 & 0.21 \\
            & 36     & C & 17 49 44.62 & $-$27 44 35.2 & $+3.4$  & 0.8 & 0.06 \\
            & 36     & C & 17 49 44.44 & $-$27 44 07.0 & $-23.2$ & 0.5 & 0.08 \\
            & 36     & E & 17 49 50.31 & $-$27 49 03.2 & $-20.1$ & 2.3 & 0.22 \\
            & 36     & E & 17 49 50.25 & $-$28 49 03.6 & $-19.9$ & 2.1 & 0.24 \\
            & 36     & E & 17 49 50.11 & $-$27 49 03.6 & $-19.4$ & 1.0 & 0.18 \\
            & 36     & E & 17 49 49.23 & $-$27 49 09.3 & $-18.6$ & 1.0 & 0.28 \\
            & 36     & E & 17 49 49.19 & $-$27 49 07.8 & $-21.9$ & 0.8 & 0.43 \\
            & 36     & E & 17 49 48.64 & $-$27 49 27.7 & $-18.1$ & 2.3 & 0.17 \\
            & 36     & E & 17 49 50.36 & $-$27 49 51.7\tablenotemark{b}& $-17.8$ & 1.8 & 0.18 \\
            & 36     & E & 17 49 51.69 & $-$27 50 12.3\tablenotemark{e}& $-17.8$ & 1.3 & 0.48 \\
G006.4$-0.1$ & 44     &   & 18 01 41.98 & $-$23 26 59.9 & $+7.0$  & 0.8 & 1.20  \\
            & 44     &   & 18 01 42.42 & $-$23 26 26.7 & $+7.8$  & 0.6 & 0.16  \\
            & 44     &   & 18 01 41.97 & $-$23 26 26.8 & $+6.8$  & 0.8 & 0.39  \\
            & 44     &   & 18 01 40.45 & $-$23 26 12.9 & $+6.8$  & 0.8 & 0.15  \\
            & 36     &   & 18 01 52.90 & $-$23 19 33.0 & $+8.2$  & 0.8 & 0.07  \\
            & 36     &   & 18 01 52.93 & $-$23 19 33.6 & $+8.2$  & 1.1 & 0.04  \\
\enddata
\label{detections}
\tablenotetext{a}{More than 100 detections were made in Sgr\,A\,East
 at both 36 GHz and 44 GHz. The detailed results of these detections are
 reported on in a separate paper by Sjouwerman et al., in
 preparation.}
\tablenotetext{b}{At the edge of the primary beam.}
\tablenotetext{c}{Contains multiple spectral features, only the brightest and most distinct ones listed.}
\tablenotetext{d}{Partly outside frequency band.}
\tablenotetext{e}{Outside the primary beam.}
\end{deluxetable*}

\section{Results}\label{results}

\subsection{Maser candidates}
Table \ref{detections} lists all detections with a signal-to-noise
ratio exceeding 10. This conservative limit was set because with
  no prior knowledge peaks with lower signal-to-noise ratios may be
  the results of random noise in such large cubes. That is, many of
  the tentative detections with lower signal-to-noise ratios were only
  made in a single channel, and thus need confirmation. Over 100
detections were made in Sgr\,A\,East, which details will be reported
separately in a forthcoming paper (see also \cite{sjouwerman10},
\cite{pihlstrom11}). In W\,28 two detections were made at 36 GHz and
four at 44 GHz. The 41 detections in G1.4$-$0.1 are all at 36 GHz. All
emission is detected as unresolved point sources, and with the single
exception of a pointing position in G1.4$-$0.1 (Sect.\ \ref{vel}) the
lines are narrow consisting of a single spectral feature. An example
of a spectral feature is plotted in Fig.\ 1. The relatively large
synthesized beam of the VLA configurations yield lower limits of the
brightness temperatures of 300$-$500 K. Such temperatures could imply
thermal emission, since the corresponding thermal linewidth would be
0.7-0.9 \kms\ and thus consistent with the widths of many of the
detections. On the other hand, a kinetic temperature of several
  hundred K is high in a molecular cloud. Molecular observations of
  W\,28 infer that the temperature of the gas region heated by the SNR
  shock is closer to 100 K, with densities of the order of 10$^5$
  cm$^{-3}$ \citep{reach05}. A test of the excitation and resulting
  flux of the 36 GHz and 44 GHz lines was made using the radiative
  transfer code RADEX \citep{vandertak07}. The available collision
  rates for methanol limits the gas temperature in the calculations to
  be below 200 K. For densities between $10^4-10^6$ cm$^{-3}$ and
  temperatures up to 200 K, only maser emission can be produced in the
  44 GHz transition. Similar results are expected for the 36 GHz line
  under such conditions, although temperatures above 20 K results in
  no convergence in the RADEX calculations. A more complete
  description of modeling of the methanol excitation conditions in
  SNRs using the radiative transfer code MOLPOP-CEP \citep{elitzur06}
  is being prepared in an accompanying paper (McEwen et al., in
  prep.), and the initial results thereof show that the 36 GHz and 44
  GHz transitions are producing maser emission under the given
  conditions.  Combining these results with the fact that several
linewidths are narrower than 0.8 \kms, and that thermal emission tends
to be widespread spatially while our detections are very compact and
spot-like, it is likely that most of the detections are masers. Future
higher resolution observations will better constrain the brightness
temperature.

\subsection{Methanol and OH spatial comparison}\label{spatial}
Our main strategy of searching for methanol masers toward OH masers
was non-optimum since the bulk of our observations resulted in
non-detections.  We discuss the reasons for this more detail in
\S\ref{detectionrate}. In a few sources, like Sgr\,A\,East, G1.4$-$0.1
and W\,28, observations included a few fields offset from the OH maser
positions. Interestingly, these three sources contain all the methanol
masers detected. In G1.4$-$0.1 a single detection is found at the edge
of the primary beam centered at the OH maser. The remaining 40 masers
are found at distances between 3-6 primary beams away from the OH
maser, corresponding to approximately 18-36 pc projected distance. In
W\,28, the two 36 GHz detections and two 44 GHz masers are located
within a primary beam distance from the nearest OH maser, while the
last two 44 GHz detections are on distances 1-1.5 beams away (1.2-1.8
pc projected distance).

\begin{figure*}[thb]\label{g1.4fig}
\resizebox{15cm}{!}{\includegraphics{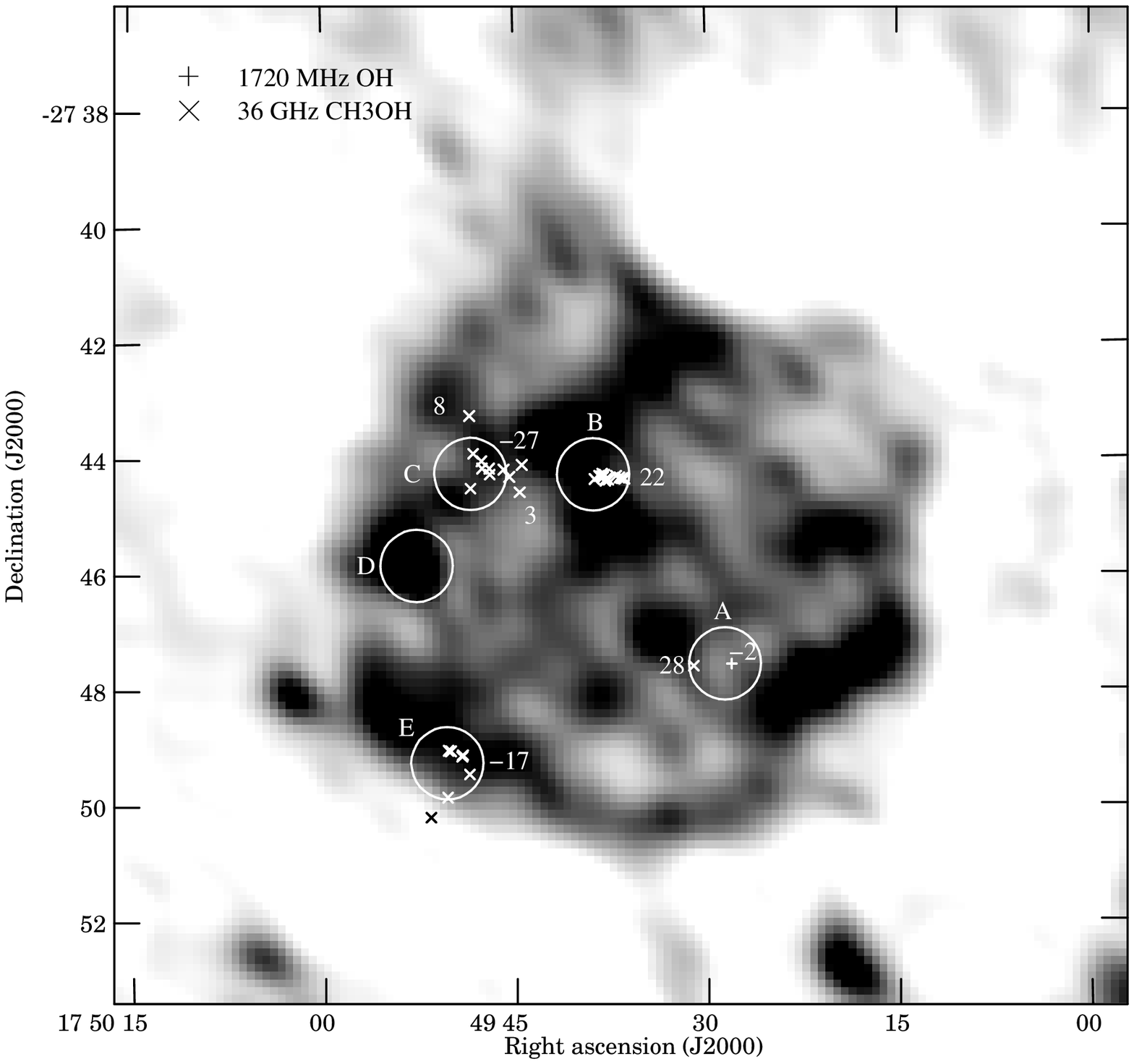}}
\caption{The 36 GHz methanol masers detected in SNR G1.4$-$0.1 are
 plotted with crosses, and the single 1720 MHz OH maser is plotted
 with a plus symbol (in A) on a grey scale 1.4 GHz continuum map of
 the SNR. Numbers indicate the mean velocities of the masers in their
 respective regions. Big circles labeled A through E show the primary
 beam field-of-view for the methanol maser search observations.}
\end{figure*}

\subsection{Methanol and OH velocity comparison}\label{vel}
The methanol maser velocities in W\,28 are similar to the OH maser
velocities at 7 \kms. This argues that the OH and methanol masers
originate from the same molecular cloud and hence are associated with
the SNR.  In contrast, the majority of the G1.4$-$0.1 masers are found
at velocities offset from the OH, at $-25$ and $+25$ \kms\ close to
the edge of the bandpass. A group of masers in G1.4$-$0.1 have
multiple peaks scattered around $+25$\kms with $\Delta V$=2$-$10
\kms. There are indications that more masers may exist at velocities
outside the reported velocity range, since some image cubes show
possible maser features near the bandpass edges that were excluded.

\begin{figure*}[thb]\label{w28full}
\rotatebox{0}{\resizebox{15cm}{!}{\includegraphics{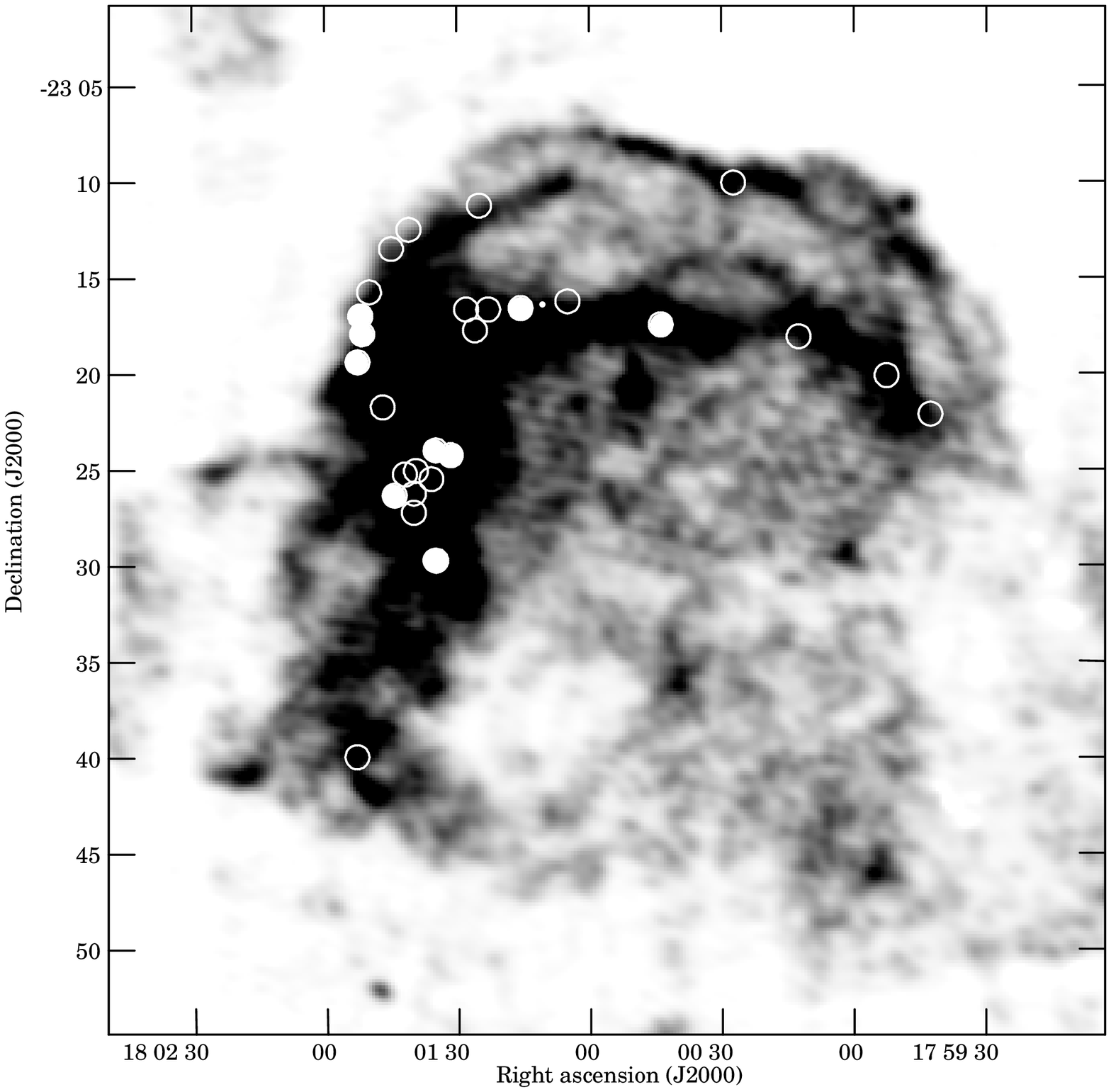}}}
\caption{Overview of the observed regions (circles) in SNR W\,28
 outlined by the 327 MHz continuum gray scale map. Regions where OH
 masers are located are marked with filled circles. The single, small
 filled circle denotes a region with an OH maser that was not
 included in the observations due to a typo in the observing
 schedule. }
\end{figure*}

\section{Discussion}\label{discussion}

\subsection{Detection rate}\label{detectionrate}
This project aims to determine whether methanol masers are a common
tracer of SNR/molecular cloud interactions, in particular compared to
the 1720~MHz OH maser tracer. Aside from Sgr\,A\,East where many
methanol masers were found, so far only two other SNRs yielded
methanol masers in the remaining 20 SNRs with OH masers. While this
absolute detection rate is not outstanding, there are two main
limitations of our search. First, the bandwidth limited the velocity
range (67 and 55 \kms\ at 36 GHz and 44~GHz, respectively) over which
we searched. This may have excluded masers if they had velocities much
offset from the observed 1720~MHz OH masers (Sect.\ \ref{vel}).

A second, and major limitation is the small field-of-view at these
high frequencies. Table \ref{targets} lists the SNR extents as
measured from low frequency (up to 2~GHz) maps and the fractional area
covered in each SNR in our observations. Except for Sgr\,A\,East and
G349.7$+$0.2 only a tiny fraction of each SNR is covered, on average
2.9\% and 1.9\% at 36~GHz and 44~GHz, respectively. A maser will
  only be found in an area actually covered by a molecular cloud, and
  the clouds presumably have a spatial covering factor $f<1$. The
  exact value of $f$ will vary from source to source and is not known
  for all individual objects. As an indicator of the size of $f$,
  observations of W\,28 suggest that about 30\% of the SNR is covered
  by dense gas \citep{atst99}.

The beam area of OH observations with the VLA is 450 to 680 times
larger than the 36 GHz or 44 GHz beam, and considering that most of
the SNRs were covered to 100\% for the OH maser searches, the number
of OH masers per unit area (assuming a cloud covering factor of
$f=0.3$) is 0.03/arcmin$^2$. Using the area actually covered by
  the methanol observations, and assuming a cloud covering factor
  $f=0.3$, the corresponding number of 36 GHz and 44~GHz masers per
  unit area is higher than the OH at 1.2 and 0.17/arcmin$^2$
  respectively.  These rough estimates exclude the detections in
Sgr\,A\,East which may be a special case. However, as will be
presented in a forthcoming paper, the number of methanol maser
detections in Sgr\,A\,East greatly exceeds the number of detected OH
masers. Therefore, we conclude that the methanol maser detection rate
is consistent with methanol masers being at least as common as the OH
masers; probably more common.

In addition to very small field-of-views, the consistent angular
offset of the methanol from the selected target OH maser positions
does not help in improving the detection rate. In Sgr\,A\,East, the
methanol masers are offset by 30-40\arcsec\ from the 1720 MHz OH
masers \citep{pihlstrom11}. Most SNRs observed in our program are less
distant and, using W\,28 as an example, such offsets translate to
angular offsets of 120-200\arcsec\ or 2-4 beams away from the OH
positions. Thus, our first order observing strategy to search for
methanol near OH masers may have actually excluded many.

\begin{figure*}[thb]\label{w28zoom}
\resizebox{15cm}{!}{\includegraphics{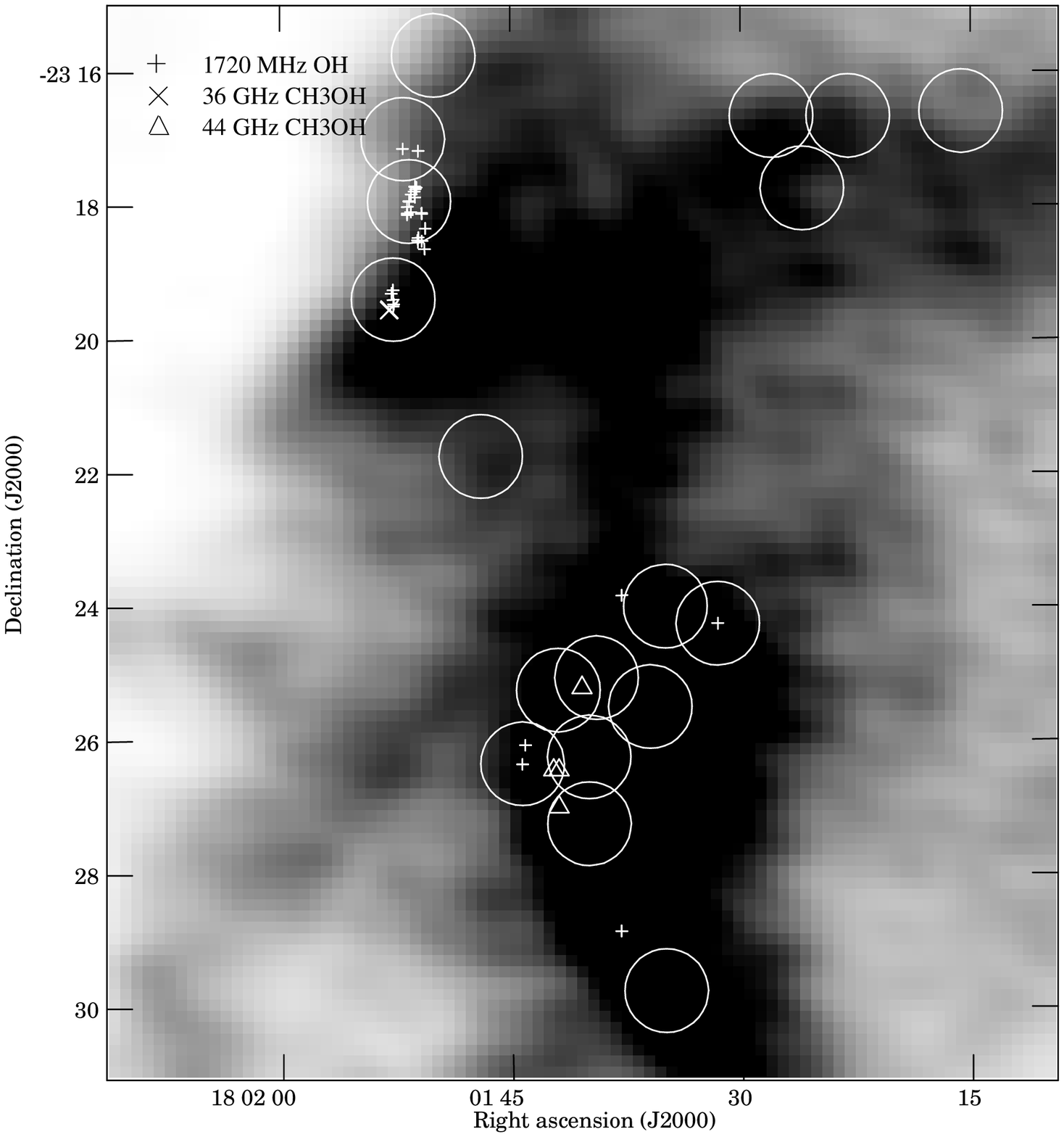}}
\caption{Both 36 GHz methanol
 masers detected in SNR W\,28 are plotted with (two overlapping)
 crosses, the four 44 GHz methanol masers with triangles
 and the 1720 MHz OH masers with plus symbols. Big circles show the
 primary beam field-of-view for the methanol
 maser search observations.}
\end{figure*}

\subsection{SNR G1.4$-$0.1}\label{g1.4}
The SNR G1.4$-$0.1 contains a large number of methanol masers. Figure
2 shows the angular distribution of the masers on a 1.4~GHz gray scale
continuum map. Less than 8\% of the SNR extent is covered by our
observations, implying that more masers may be associated with this
object. The mean velocities of the masers at each pointing position,
indicated in Figure 2, differs significantly from the
velocity of the single 1720 MHz OH maser at $+2$ \kms. The methanol is
found over the full 60 \kms\ velocity range observed, from about
$-$30 \kms\ to $+$30 \kms. Since many masers were detected close to
either edge of that range, it is possible that additional masers at
higher negative and positive velocities exist.

G1.4$-$0.1 has not been studied in much detail as an individual object
except for the radio data defining it as an SNR
\citep{liszt92,gray94,bhatnagar02} and the detection of a 1720 MHz OH
maser by \citet{yusef-zadeh99}. However, due to its proximity to the
Galactic Center the SNR position is covered in a number of surveys
toward the GC and coincides with several different species of
molecular emission. \citet{oka98} performed a survey of $^{12}$CO and
$^{13}$CO of the Galactic Center region including the coordinates of
G1.4$-$0.1. Both CO isotopes are detected toward the position of
G1.4$-$0.1, and this region is noted by the authors as belonging to a
feature labeled the 'molecular flare'. This flare includes molecular
velocities from $+30$ to $+140$ \kms, and consists of both filaments
and 'fluffy clouds'. The linewidths are large, consistent with
large-scale shocks. \citet{oka98} also point out that this region
contains arcs and shells of 10-20 pc diameter, implying a dynamic and
complex area on the sky. Similarly, \citet{tsuboi99} report on CS(1-0)
emission in this region, where there is a gap in the CS velocity
coverage around 0 \kms, and two prominent velocity features at $-$40
and $+$40\kms. In their data they define a more distinct feature,
where much of the emission belongs to the end of a molecular
ridge/streamer arcing in toward the central region of the Galaxy.
Comparing these molecular maps to our methanol data, it seems as if
G1.4$-$0.1 is interacting with two molecular clouds, or with two
individual velocity features belonging to the molecular streamer. The
negative velocity methanol masers observed at the Eastern side of the
SNR agree well with the $-$40 \kms\ cloud, and similarly, the positive
velocity methanol masers at the Western side agree with the $+$40
\kms\ molecular cloud.

One question to address is whether the methanol masers are associated
with an interaction between the SNR and the two molecular clouds
observed in CO and CS along the line-of-sight. While the presence of
an OH 1720 MHz OH maser normally is taken as a signpost of the passage
of a C-type shock \citep{lockett99}, the OH maser velocity in
G1.4$-$0.1 is offset by 5--30\kms\, from the methanol maser
velocities. This could perhaps imply that the methanol masers are not
associated with the same SNR/molecular cloud interaction. Other
possibilities could be like the one explored by \citep{yusef-zadeh13},
where bright 36 GHz methanol maser candidates detected in the inner GC
region are explained by an enhanced methanol abundance due to cosmic
ray induced photodesorption in molecular clouds. The pumping mechanism
of these maser candidates is not clear, but could be from local shocks
or, less likely, due to radiation from nearby star formation. However,
the offset in velocity between the OH and methanol alone does not
exclude an interaction origin. In particular, a similar velocity
offset of 5--30\kms\, between the 36 GHz methanol and 1720 MHz OH
masers in Sgr\,A\,East is observed \citep{sjouwerman10}, and may even
be a natural consequence of the methanol masers originating closer to
the shock front while the OH masers are produced in the post-shock
gas. Blue-shifted OH may originate in swept-up gas coming toward us,
while methanol is found deeper in the cloud. Other evidence for an
interaction may come from morphology of the molecular clouds. Studying
the CS and CO channel maps presented by \citet{tsuboi99} and
\citet{oka98}, there appears to be a cavity around the extent of the
SNR (as outlined by radio continuum), especially significant for the
higher velocity cloud on the western side. At the negative velocities,
there appears to be an abrupt morphological boundary at the
southeastern edge of the remnant, but not as much encompassing
morphologically. These morphological signatures (at least for the
higher velocities) are remarkably similar to what has been observed in
$^{12}$CO and $^{13}$CO in the SNR 3C397 \citep{jiang10}, where a
molecular cavity is observed around the SNR. The case for a SNR/cloud
interaction in 3C397 is strengthened by a broadening of the blueward
side of the spectra toward the interaction region. Comparable
kinematical spectral data for G1.4$-$0.1 is not readily available in
the CS and CO surveys, and it is difficult to make such a detailed
study based on the longitude/latitude-velocity plots alone. Additional
SNR/cloud interaction evidence may come from emission from shock
excited H$_2$ (1-0) S(1), or from line ratios showing a high
excitation (e.g., a high CO(2-1)/CO(1-0)) ratio. None of these
particular pieces of information currently exists for G1.4$-$0.1, but
will be explored in the future. However, existing survey data of the
Galactic Center region exist in SiO, HCO+ and H$^{13}$CO+
\citep{riquelme10}. From this data we find that the SiO which is often
associated with shocks, peaks at +60 \kms\ and around $-$20 \kms\,
across G1.4$-$0.1 (extending through the velocities down to +0 \kms\
covered by our methanol search). A similar behaviour is seen also in
HCO+ and H$^{13}$CO+. The presence of shocks is supported by the
elevated SiO/HCO+ ratio at the position of G1.4$-$0.1 at both cloud
velocities. By studying the cubes a signature of a cavity can also be
seen like in the CO maps. Again, this signature is much more obvious
at the higher velocity cloud.

While there is no conclusive evidence for a SNR/cloud interaction for
the clouds where the methanol originates, there is also no evidence
for star formation in these clouds causing enhanced methanol abundance
or providing pumping. Given the molecular morphology and the
supportive evidence of shocks via the OH, SiO and HCO+, an interaction
scenario currently seems to best explain the presence of methanol
masers.

The positive velocity masers in G1.4$-$0.1 (position B in Fig.\ 2)
differ in their characteristics compared to the other masers
detected. This group of masers show multiple spectral features,
covering a range of velocities. They further align in the East-West
direction, which may be a geometry effect, perhaps outlining a region
parallel to a shock front where long path lengths of velocity coherent
gas exists. Such a shock front may be associated with some of the gas
observed in the CO molecular flare region.

All of the masers in G1.4$-$0.1 are observed in the 36~GHz
transition. Preliminary results from methanol maser modeling (McEwen
et al., in prep.) show that the 36 GHz transition dominates at high
gas densities. This would imply that the gas may be associated with a
shock front, in agreement with what was found in the GC
\citep{sjouwerman10}. Given the broad overlap in physical conditions
under which 36 GHz and 44 GHz methanol masers arise, it is intriguing
that 44 GHz masers are not detected in this SNR that is otherwise rich
in 36 GHz masers. A possible explanation could be that 44 GHz emission
only appears in regions of these molecular clouds at velocities beyond
the range observed here.

\subsection{SNR W\,28}

W\,28 is the archetype of an SNR interacting with a molecular cloud.
The evidence for such an interaction for W\,28 comes from several
sources.  It is the first and best-studied SNR to have larger numbers
of shock-excited OH 1720 MHz masers
\citep{fgs94,claussen97,cgfd99,hgbc05}. Observations of millimeter and
near-infrared lines reveal bright clumps of dense molecular gas with
broad-line profiles indicative of shock disruption
\citep{atst99,reach05,nrb+11,nrb+12}. Figure 3 overlays the observed
primary beams on a grey scale 327 MHz continuum map of SNR W\,28. The
beams with OH masers are indicated as filled circles, other observed
beams are open circles. A zoom of the Eastern side is shown in Fig.\
4, plotting the relative positions of the OH and methanol masers. The
methanol masers are all detected within a tight velocity range between
6.8$-$8.2 \kms, with no spread or multiple lines. The close agreement
between the velocity of the methanol masers and both the molecular
cloud and the OH masers, argues for a physical association rather than
a line-of-sight coincidence.

All the masers are found in the northeastern cloud that is thought to
delineate a strong interaction of W\,28 SNR with a molecular cloud.
However, the positions of 44 GHz masers do not not coincide with any
36 GHz masers, implying a density or temperature gradient between the
regions. The 36 GHz masers are coincide with the broad line regions
defined by \citet{reach05} at the northern part of this cloud. The 44
GHz masers are found near the peak of Core 2, a region rich with
shocked molecular species including NH$_3$, CS, SiO and thermal
CH$_3$OH, detected in the extensive 7 mm and 12 mm surveys by
\citet{nrb+11,nrb+12}. Of particular interest is this northeastern
cloud is also the location of the TeV source HESS J1801$-$233
\citep{aab+08}. Both the Fermi and AGILE satellites have detected GeV
emission in this vicinity also \citep{aaa+10,gtb+10}. The TeV emission
is found extending across both groups of masers, but is seems to be
stronger near the group of 44 GHz masers. W\,28 is thus an interesting
source to test hadronic cosmic ray models using masers. This study
will be deferred to a future paper.

\subsection{Methanol as an SNR/molecular cloud tracer}

The formation of methanol molecules is enhanced in regions where icy
dust grains are subject to UV rays, cosmic ray irradiation or some
other heating mechanism \citep{whittet11}. In SNRs no UV irradiation
is expected, but in regions where an SNR interacts with the
surrounding molecular gas, cosmic rays may be produced and
accelerated. Cosmic rays have been shown to be a viable source of
heating resulting in widespread methanol maser emission in molecular
clouds throughout the inner region of the Galaxy
\citep{yusef-zadeh13}. Thus, the interaction between the SNR and the
molecular cloud provides a direct source of heating via slow and
non-dissociative shocks. In W\,28 and G1.4$-$0.1 the OH masers
indicate such interactions. In G1.4$-$0.1 there are at least two
distinct molecular clouds interacting with the SNR shell, which
probably explains the wealth of masers in this source. The molecular
cloud associated with W\,28 is displaying less turbulent features,
consistent with a more constrained interaction region.

These observations show that Class I methanol masers exist in other
SNRs than Sgr\,A\,East, and that they are associated with 
molecular cloud interactions. As such, they are useful line tracers
to estimate physical conditions in the interaction
regions. It is not yet clear whether they are more abundant than the OH
masers due to the limited angular and frequency coverage of our
observations. Complete surveys of SNRs are expensive at these
frequencies due to the small primary beams.

In Sgr\,A\,East we have previously argued that the bright 36~GHz masers are
better tracers of the shock front than the 44 GHz masers
\citep{pihlstrom11}. Ongoing modeling of methanol maser conditions
implies that this situation occurs at relatively high densities, consistent
with a special region of the shock front and at higher densities than
where the OH is found (McEwen et al., in prep.). The 44 GHz masers may
be found deeper in the cloud, where the gas yet has to be compressed
and heated by the shock front. With slightly different excitation
conditions there may also be a difference in geometry, since bright
masers arise along specific, long path lengths of velocity coherent
gas. More detailed modeling and comparison of conditions giving rise to the
different maser lines will be discussed elsewhere.

\section{Summary}
We have presented the results of a targeted search for Class I
methanol masers in a sample of Galactic SNRs with known 1720 MHz OH
masers, and thus a search in SNRs that interact with the surrounding
molecular environment. Due to limited angular coverage and a
restricted velocity range, the methanol detections were limited to
three SNRs: Sgr\,A\,East, W\,28 and G1.4$-$0.1. We conclude that at
this point information is insufficient to determine whether methanol
masers are better SNR/MC interaction tracers than OH masers, though
the brightest 36 GHz methanol maser may trace the denser shock front
the most accurate. Surveys covering a larger part of the SNRs are
needed or better estimates for the location of methanol (i.e., other
than the OH masers) must be determined. Other future work includes
radiative transfer modeling of methanol, and applying the maser
detections in Sgr\,A\,East, W\,28 and G1.4$-$0.1 to study the hadronic
cosmic ray model.

\acknowledgments We would like to acknowledge D.\,Riquelme for kindly
providing the SiO, HCO+ and H$^{13}$CO data cubes. Y.M.P.\,and
B.M.\,were partly supported by NASA grant NNX10A055G. The National
Radio Astronomy Observatory is a facility of the National Science
Foundation operated under cooperative agreement by Associated
Universities, Inc.

{\it Facilities:} \facility{VLA}, \facility{FERMI}.


\begin{thebibliography}{}

\bibitem[Aharonian et al.(2008)]{aab+08} Aharonian, F., Akhperjanian,
A.~G., Bazer-Bachi, A.~R., et al.\ 2008, \aap, 481, 401

\bibitem[Abdo et al.(2010)]{aaa+10} Abdo, A.~A., Ackermann,
M., Ajello, M., et al.\ 2010, \apj, 718, 348

\bibitem[Arikawa et al.(1999)]{atst99} Arikawa, Y., Tatematsu,
K., Sekimoto, Y., \& Takahashi, T.\ 1999, \pasj, 51, L7

\bibitem[Bhatnagar, S.(2002)]{bhatnagar02}Bhatnagar, S., 2002, MNRAS,
 332, 1

\bibitem[Brogan et al.(2013)]{brogan13}Brogan, C.L., Goss, W.M.,
  Hunter, T.R., et al., 2013, ApJ, 771, 91

\bibitem[Brogan et al.(2000)]{brogan00}Brogan, C.~L., Frail, D.A.,
 Goss, W.~M., \& Troland, T.~H., 2000, \apj, 537, 875

\bibitem[Claussen  et  al.(1997)]{claussen97}Claussen,  M.~J.,  Frail,
 D.~A., Goss, W.~M., \& Gaume, R.~A., 1997, \apj, 489, 143

\bibitem[Claussen et al.(1999)]{cgfd99} Claussen, M.~J., Goss,
W.~M., Frail, D.~A., \& Desai, K.\ 1999, \apj, 522, 349

\bibitem[Cragg et al.(2002)]{cragg02}Cragg, D.M., Sobolev, A.M., \&
  Godfrey, P.D., 2002, MNRAS, 331, 521

\bibitem[Cragg et al.(1992)]{cragg92}Cragg, D.~M., Johns, K.~P.,
 Godfrey, P.~D., \& Brown, R.~D., 1992, \mnras, 259, 203

\bibitem[Cristofari et al.(2013)]{cristofari13}Cristofari, P., Gabici,
  S., Casanova, S., Terrier, R., \& Parizot, E., 2013, MNRAS, 434,
  2748

\bibitem[Drury et al.(1994)]{drury94}Drury, L.~O'C., Aharonian, F.~A.,
 \& V\"olk, H.~J., 1994, \aa, 287, 959

\bibitem[Elitzur \& Asensio Ramos(2006)]{elitzur06}Elitzur, M., \&
  Asensio Ramos, A., 2006, MNRAS, 365, 779

\bibitem[Elitzur(1976)]{elitzur76}Elitzur, M., 1976, ApJ, 203, 124

\bibitem[Fish et al.(2011)]{fish11}Fish, V.L., Muehlbrad, T.C.,
  Pratap, P., et al., 2011, ApJ, 729, 14

\bibitem[Frail \& Mitchell(1998)]{frail98}Frail, D.A., \& Mitchell,
G.F., 1998, ApJ, 508, 690

\bibitem[Frail et al.(1994)]{fgs94} Frail, D.~A., Goss, W.~M., \&
Slysh, V.~I.\ 1994, \apjl, 424, L111

\bibitem[Frail(2011)]{frail11} Frail, D.~A.\ 2011, \memsai, 82, 703

\bibitem[Giuliani et al.(2010)]{gtb+10} Giuliani, A.,
Tavani, M., Bulgarelli, A., et al.\ 2010, \aap, 516, L11

\bibitem[Gray(1994)]{gray94}Gray, A.D., 1994, MNRAS, 270, 847

\bibitem[Green et al.(1997)]{green97}Green, A.~J., Frail, D.~A.,
   Goss, W.~M., \& Otrupcek, R., 1997, \aj, 114, 2058

\bibitem[Green(2009)]{green09}Green D. A., 2009, Bulletin of the
   Astronomical Society of India, 37, 45

\bibitem[Hewitt et al.(2008)]{hewitt08}Hewitt, J.~W., Yusef-Zadeh,
F., \& Wardle, M., 2008, \apj, 683, 189

\bibitem[Hewitt et al.(2009)]{hewitt09}Hewitt, J~W., Yusef-Zadeh,
   F., \& Wardle, M., 2009, \apjl, 706, L270

\bibitem[Hoffman et al.(2005)]{hgbc05} Hoffman, I.~M., Goss,
W.~M., Brogan, C.~L., \& Claussen, M.~J.\ 2005, \apj, 620, 257

\bibitem[Jiang et al.(2010)]{jiang10} Jiang, B., Chen, Y., Wang, J.,
  et al.\ 2010, \apj, 712, 1147


\bibitem[Kalenskii et al.(2010)]{kalenskii10}Kalenskii, S.V.,
  Johansson, L.E.B., Bergman, P., et al., 2010, \mnras, 405, 613

\bibitem[Karlsson et al.(2003)]{karlsson03}Karlsson, R., Sjouwerman,
  L.~O., Sandqvist, Aa., \& Whiteoak, J.~B., 2003, \aap, 403, 1011

\bibitem[Kelner et al.(2006)]{kelner06}Kelner, S.R., Aharonian, F.A.,
  \& Bugayov, V.V., 2006, PhRvD, 74, 4018

\bibitem[Koralesky et al.(1998)]{koralesky98}Koralesky, B., Frail,
 D.~A., Goss, W.~M., Claussen, M.~J., \& Green, A.~J., 1998, \aj,
 116, 1323

\bibitem[Larionov \& Val'tts(2007)]{larionov07}Larionov, G.M., \&
  Val'tts, I.E., 2007, ARep, 51, 756

\bibitem[Lazendic et al.(2010)]{lazendic10}Lazendic, J.~S., Wardle,
  M., Whiteoak, J.~B., Burton, M.~G., \& Green, A.~J., 2010, \mnras,
  409, 371

\bibitem[Leurini et al.(2007)]{leurini07}Leurini, S., Schilke, P.,
  Wyrowski, F., \& Menten, K.M., 2007, \aap, 466, 215

\bibitem[Liszt(1992)]{liszt92}Liszt, H., 1992, ApJSS, 82, 495

\bibitem[Litovchenko et al.(2012)]{litovchenko12}Litovchenko, I.D.,
  Bayandina, O.S., Alakoz, A.V., et al., 2012, ARep, 56, 536

\bibitem[Lockett et al.(1999)]{lockett99}Lockett, P., Gauthier, E., \&
 Elitzur, M., 1999, \apj, 511, 235

\bibitem[Morimoto et al.(1985)]{morimoto85}Morimoto, M., Kanzawa, T.,
 \& Ohishi, M., 1985, \apjl, 288, L11

\bibitem[Nicholas et al.(2012)]{nrb+12} Nicholas, B.~P.,
Rowell, G., Burton, M.~G., et al.\ 2012, \mnras, 419, 251

\bibitem[Nicholas et al.(2011)]{nrb+11} Nicholas, B., Rowell,
G., Burton, M.~G., et al.\ 2011, \mnras, 411, 1367

\bibitem[Oka et al.(1998)]{oka98}Oka, T., Hasegawa, T., Sato, F.,
 Tsuboi, M., \& Miyasaki, A., 1998, ApJSS, 118, 455

\bibitem[Pierce-Price et al.(2000)]{pierce-price00}Pierce-Price, D.,
 et al., 2000, ApJL, 545, L121

\bibitem[Pihlstr\"om et al.(2011)]{pihlstrom11}Pihlstr\"om, Y.~M.,
 Sjouwerman, L.~O., \& Fish, V.~L., 2011, \apjl, 739, L21

\bibitem[Pratap et al.(2008)]{pratap08}Pratap, P., Shute, P.~A.,
 Keane, T.~C., Battersby, C., \& Sterling, S., 2008, \aj, 135, 1718

\bibitem[Reach et al.(2005)]{reach05}Reach, W.T., Rho, J., \& Jarrett,
 T.H., 2005, \apj, 618, 297

\bibitem[Riquelme et al.(2010)]{riquelme10}Riquelme, D., Bronfman, L.,
 Mauersberger, R., May, J., \& Wilson, T.L., 2010, A\&A, 523, A45

\bibitem[Sjouwerman et al.(2010)]{sjouwerman10}Sjouwerman, L.~O.,
 Pihlstr\"om, Y.~M., \& Fish, V.~L., 2010, \apjl, 710, L111

\bibitem[van der Tak et al.(2007)]{vandertak07}van der Tak, F.~F.~S.,
  Black, J.~H., Sch\"oier, F.~L., Jansen, D.~J., \& van Dishoeck,
  E.~F., 2007, \aap, 468, 627

\bibitem[Tsuboi et al.(1999)]{tsuboi99}Tsuboi, M., Handa, T., \&
 Ukita, N., 1999, ApJSS, 120, 1

\bibitem[Wardle \& Yusef-Zadeh(2002)]{wardle02}Wardle, M. \&
 Yusef-Zadeh, F., 2002, Science, 296, 2350

\bibitem[Wardle(1999)]{wardle99}Wardle, M., 1999, ApJ, 525, L101

\bibitem[Whittet et al.(2011)]{whittet11}Whittet, D.C.B., Cook, A.M.,
 Herbst, E., Chiar, J.E., \& Shenoy, S.S., 2011, \apj, 742, 28

\bibitem[Yusef-Zadeh et al.(1999)]{yusef-zadeh99}Yusef-Zadeh, F.,
 Goss, W.M., Roberts, D.A., Robinson, B., \& Frail, D.A., 1999, \apj,
 527, 172

\bibitem[Yusef-Zadeh et al.(2008)]{yusef-zadeh08}Yusef-Zadeh, F.,
 Braatz, J., Wardle, M., \& Roberts, D., 2008, \apjl, 683, L147

\bibitem[Yusef-Zadeh et al.(2013)]{yusef-zadeh13}Yusef-Zadeh, F.,
 Cotton, W., Viti, S., Wardle, M., \& Royster, M., 2013, ApJL, 764,
 L19

\end{thebibliography}
\end{document}